\documentclass[twocolumn,iop]{emulateapj}
\usepackage{natbib}

\newcommand{\afe}{[$\alpha$/Fe]}

\begin{document}

\title{Evidence for the rapid formation of low mass early-type galaxies\\ 
         in dense environments }

\author{Yiqing Liu\altaffilmark{1,2,3}, Eric W. Peng\altaffilmark{1,2,3}, John Blakeslee\altaffilmark{4}, Patrick C{\^o}t{\'e}\altaffilmark{4}, Laura Ferrarese\altaffilmark{4}, Andr\'es Jord\'an\altaffilmark{5}, Thomas H.\ Puzia\altaffilmark{5}, Elisa Toloba\altaffilmark{9,10}, Hong-Xin Zhang\altaffilmark{6,7,8,1,2}}

\altaffiltext{1}{Department of Astronomy, Peking University, Beijing 100871, China.}
\altaffiltext{2}{Kavli Institute for Astronomy and Astrophysics, Peking University, Beijing 100871, China}
\altaffiltext{3}{Corresponding authors: yiqing.liu@pku.edu.cn, peng@pku.edu.cn}
\altaffiltext{4}{Herzberg Institute of Astrophysics, National Research Council of Canada, Victoria, BC V9E 2E7, Canada}
\altaffiltext{5}{Instituto de Astrof\'isica, Facultad de F\'isica, Pontificia Universidad Cat\'olica de Chile, Av. Vicu\~{n}a Mackenna 4860, 7820436 Macul, Santiago, Chile}
\altaffiltext{6}{National Astronomical Observatories, Chinese Academy of Sciences, A20 Datun Rd, Chaoyang District, Beijing 100012, China}
\altaffiltext{7}{CAS-CONICYT Fellow}
\altaffiltext{8}{Chinese Academy of Sciences South America Center for Astronomy, Camino EI Observatorio \#1515, Las Condes, Santiago, Chile}
\altaffiltext{9}{UCO/Lick Observatory, University of California, Santa Cruz, 1156 High Street, Santa Cruz, CA 95064, USA}
\altaffiltext{10}{Texas Tech University, Physics Department, Box 41051, Lubbock, TX 79409-1051, USA}

\begin{abstract}

We explore the environmental dependence of star formation timescales in low mass galaxies using the \afe\ abundance ratio as an evolutionary clock. We present integrated \afe\ measurements for 11 low mass ($M_\star \sim 10^9~M_\odot$) early-type galaxies (ETGs) with a large range of cluster-centric distance in the Virgo Cluster. 
We find a gradient in [$\alpha$/Fe], where the galaxies closest to the cluster center (the cD galaxy, M87) have the highest values. This trend is driven by galaxies within a projected radius of 0.4~Mpc (0.26 times the virial radius of Virgo~A), all of which have super-solar [$\alpha$/Fe]. Galaxies in this mass range exhibit a large scatter in the \afe--$\sigma$ diagram, and do not obviously lie on an extension of the relation defined by massive ETGs. In addition, we find a correlation between [$\alpha$/Fe] and globular cluster specific frequency ($S_N$), suggesting that low-mass ETGs that formed their stars over a short period of time, were also efficient at forming massive star clusters. 
The innermost low-mass ETGs in our sample have \afe\ values comparable to that of M87, implying that environment is the controlling factor for star formation timescales in dense regions. These low-mass galaxies could be the surviving counterparts of the objects that have already been accreted into the halo of M87, and may be the link between present-day low-mass galaxies and the old, metal-poor, high-\afe, high-$S_N$ stellar populations seen in the outer halos of massive ETGs.

\end{abstract}



\section{Introduction}

Environment plays a critical role in the evolution of low mass galaxies. 
Their relatively shallow potential wells make them most vulnerable to external influences, which can affect their gas content, star formation history, morphology, and kinematics. 
For instance, the fraction of quenched galaxies with masses below $10^9~M_\odot$ depends strongly on the distance to a nearby massive companion \citep{vdBergh_94,Geha_12}. Unlike their massive early-type counterparts, whose spheroidal shapes are believed to be a result of mergers, low-mass early-type galaxies (ETGs) are less likely to have experienced major mergers in the cluster environment because of their high relative velocities. Non-merger processes, however, are also able to quench and transform low-mass ETGs (i.e. "dwarf ETGs", which is more commonly used; M$_B \gtrsim-$18). Environmental processes such as ram pressure stripping \citep{Gunn_Gott_72}, harassment \citep{Moore_96}, and tidal stirring \citep{Mayer_01} all lead to lower gas fractions and spheroidal morphology. 
In the simulations of \citet{Mastropietro_05}, a significant fraction of a late-type galaxy's disk is able to survive when it falls into galaxy clusters. 
From observations, more and more low-mass ETGs are found to have complex fine structures in both morphology (such as arms and disks, e.g., \citealt{Jerjen_00,Barazza_02,Geha_03,Graham_Guzman_03,deRijcke_03,Lisker_06,Ferrarese_06,Janz_12,Janz_14,Guerou_15}) and kinematics (such as rotating or counter-rotating, e.g. \citealt{Pedraz_02,Simien_02,Geha_02,Geha_03,vanZee_04,Chilingarian_09,Toloba_09,Toloba_11,Rys_13,Rys_14,Toloba_14_II,Guerou_15,Toloba_15}) that are more characteristic of disk galaxies. 
In addition, \citet{Lisker_06} measured the fraction of light in spiral arm substructures, taking into account their pitch angles as well, and concluded that low mass ETGs with spiral structure truly are disk galaxies rather than just having a small disk component. 
Therefore, there is evidence that today's dwarf ETGs may have had disk-dominated, gas-rich progenitors, and that the transformation could have been driven by environmental processes.

There is, however, also a different kind of environmental influence on low mass galaxies, one that is intrinsic to the galaxy. $\Lambda$CDM simulations of large-scale structure show that dark matter halos with lower mass collapse first, and those ones located in denser environments collapse earlier (e.g., \citealt{Gao_05}). Therefore, in the $\Lambda$CDM picture, low mass galaxies in dense environments formed in different kinds of dark matter halos---these halos collapsed earlier, perhaps allowing an earlier start to star formation, and they were also denser. Semi-analytic models of galaxy formation show that low mass galaxies in denser regions have a narrower peak star formation rate (SFR), and the times at which the SFR peaked are earlier than those for their counterparts in less dense regions (\citealt{deLucia_06}; \citealt{P08}). Combining these results leads to an expectation that dwarf galaxies in denser regions had more intense star formation at earlier times.

A good place to study environmental effects, both ``internal'' and ``external'', is in galaxy clusters. A galaxy cluster typically samples a wide range of mass density, from the cluster center to its outskirts. Galaxy clusters also harbor a large fraction of the quenched (early-type) galaxies in the Universe. The Virgo Cluster, as the nearest galaxy cluster ($D\sim$16.5~Mpc), provides us with our best opportunity to study a large number of faint, low mass ETGs across a range of galaxy density. 

One clue that low mass Virgo ETGs had environmentally-influenced star formation histories is from their globular cluster (GC) systems. The GC specific frequency ($S_N$, the ratio of the number of GCs to stellar $V$-band luminosity) is an oft-used diagnostic to compare a galaxy's star cluster and star formation histories. In local star forming galaxies, the mass of the most massive young star cluster is directly correlated with the current star formation rate \citep{Weidner_04,Bastian_08}. Because GCs are often the most massive star clusters produced in a galaxy's evolution (with nuclear clusters being the exception), the number of GCs is a rough proxy for the fraction of a galaxy's star formation that occurred above a certain (high) threshold SFR. In the denominator of GC specific frequency is the total stellar luminosity (or mass) of a galaxy, which represents the time-averaged SFR. In this context, $S_N$ is a measure of how ``peaked'' a galaxy's star formation history was. 

Using data from the ACS Virgo Cluster Survey (ACSVCS; \citeauthor{Cote_04} 2004), \cite{P08} found that dwarf ETGs closer to the dynamical center of the Virgo A subcluster (the cD galaxy M87) had higher GC specific frequencies. They interpreted this as evidence that the dwarf ETGs in the dense, central regions, had a more intense early star formation history that formed more massive star clusters, i.e., they had higher peak SFRs, forming more of their stars over a shorter period of time.

There is a more direct way to test this idea, however, and that is to measure the star formation timescales of the stars themselves. The intensity of star formation and the rapidity of quenching in galaxies are imprinted on the age, metallicity, and abundance distributions of the stars. While obtaining star-by-star formation histories for these galaxies is currently beyond our capability, integrated spectra can still provide clues to the history of the dominant stellar population. In particular, the alpha element-to-iron abundance ratio, [$\alpha$/Fe], is a reliable indicator of early star formation intensity. After star formation begins, high-mass stars evolve the quickest and return their gas and metals to the ISM as Type~II supernovae (SNe II). After $\sim0.1$ Gyr \citep{Maoz_10}, Type~Ia supernova (SNe Ia), which evolve from low mass stars, begin to explode and return their gas and metals to the ISM. Because SNe II ejecta have a higher [$\alpha$/Fe] than that of SNe Ia, the mean [$\alpha$/Fe] of the galaxy stays high at early times, but then decreases as SNe~Ia appear. As a result, galaxies that finish their star formation quickly (more rapid star formation or earlier quenching, i.e., a shorter star formation timescale) are expected to have higher [$\alpha$/Fe]. Measuring [$\alpha$/Fe] of low mass ETGs tells us about these galaxies' average star formation timescale. Moreover, if we do this for many galaxies at different radii of the Virgo Cluster, we have a way to test whether environment significantly affects the star formation timescale in progenitors of today's low mass ETGs. 

Although this is a promising technique in principle, there are many observational obstacles. The low surface brightness ($\mu_B\gtrsim$22~mag/arcsec$^2$) of these low mass ETGs makes it difficult to obtain spectra with signal-to-noise ($S/N$) high enough to do stellar population analyses. Also, the central regions of dwarf ETGs often host a nuclear star cluster whose stellar populations may not be indicative of the galaxy as a whole. These bright nuclei are typically younger and more metal-rich than the main stellar bodies of galaxies, and they can contribute a non-negligible amount of light in the central regions, where long slit spectra are typically taken.

Nevertheless, there have been a few earlier studies, but only for the brightest dwarfs ($M_r \lesssim -15.5$). \citet{Michielsen_08} found no difference in [$\alpha$/Fe] between dwarf ETGs located in the field and the Virgo Cluster, with measured values all scattered around solar. Similarly, using a Virgo sample with a similar range of age and metallicity, \citet{Paudel_10_dE} found no relation between $[\alpha/Fe]$ and local densities. These studies did not exclude nuclear star clusters from their galaxy spectra, probably due to insufficient $S/N$ in the outer regions of their slits, making interpreting these results difficult.

Our solution to these problems is to use IFU (integral field unit) spectroscopy. Using an IFU we can observe a larger amount of area in the inner regions, where the surface brightness is relatively higher, and we can exclude spectra taken of the nucleus. By combining all the spectra in these regions, we can obtain sufficient signal to study fainter galaxies.

In this work, we present [$\alpha$/Fe] measurements for the main stellar bodies (excluding the nuclear light) of 11 Virgo low mass ETGs with a large range of cluster-centric distance. The paper is structured as follows: $\S$\ref{data} describes sample selection, observation, and data reduction. $\S$\ref{ssp} and $\S$\ref{results} describe data analysis and results, respectively. We discuss the application of our results and the relation with globular cluster formation in $\S$\ref{discussion}, and list the main conclusions in $\S$\ref{conclusion}.

\section{Data}
\label{data}

\subsection{Sample}

We used the HST/ACS Virgo Cluster Survey (ACSVCS; \citeauthor{Cote_04} 2004) as our parent sample. The ACSVCS imaged 100 Virgo Cluster early-type galaxies (with $B_T<16$~mag). For these 100 galaxies, the survey provides us with structural parameters and photometry \citep{Ferrarese_06}, nuclear properties \citep{Cote_06}, and globular cluster color distributions \citep{Peng_06}, luminosity functions \citep{Jordan_07}, and specific frequencies \citep{P08}. The specific frequency ($S_N$) of globular clusters is defined as the number of globular clusters (GCs) per unit of galactic luminosity, normalized to $M_V=-15$ \citep{Harris_vdB_81}. 
\citet{P08} found that for ACSVCS galaxies fainter than $M_z = -19$, $S_N$ has an environmental dependence. Those dwarf galaxies that are closer to M87 typically have higher $S_N$. To explore the origin of this environmental dependence, we selected 11 ETGs in the faintest 1.5 mag of the ACSVCS sample ($-17.5 < M_V < -16$) with the aim of measuring their [$\alpha$/Fe]. These galaxies cover a large range of $S_N$ and cluster-centric distance (six are within 0.4~Mpc projected radius ($R_p$) from M87 and five have $0.4<R_p<1.5$~Mpc). The basic parameters of our sample are listed in Table~\ref{dE_param}. 

In Table~\ref{dE_param}, velocity dispersions ($\sigma_e$) for our sample galaxies are included. For four of these galaxies (VCC~230, 1185, 1539, and 1545), we present previously unpublished measurements. These measurements were made using spectra taken with the Keck/ESI spectrograph \citep{Sheinis_02} in February, 2003, and March, 2004. Exposure times of them varied from 2700s to 5400s, and the slit width was $0\farcs75$ or $1\farcs0$, depending on the seeing conditions. The spectral dispersion was $11\ {\rm km\ s^{-1}\ pixel^{-1}}$. Line-of-sight integrated velocity dispersions were measured by broadening stellar templates, as described in Section~4 of \citep{Hasegan_05}. The nuclei were excluded when summing up the light along the slit.

\newcommand\tna{\,\tablenotemark{a}}
\begin{deluxetable*}{cccccccccc}
\tabletypesize{\scriptsize}
\tablecaption{Properties of the low mass ETGs in our sample. \label{dE_param}}
\tablewidth{0pt}
\tablehead{
\colhead{VCC} & 
\colhead{RA} & \colhead{Dec} &
\colhead{$\sigma_e$} &
\colhead{$M_V$} & \colhead{$M_z$} &
\colhead{$M_\star$} & \colhead{$R_p$} & 
\colhead{$S_N$} & \colhead{$S_{N,z}$} \\
\colhead{} & 
\colhead{(J2000)} & \colhead{(J2000)} &
\colhead{(km/s)} &
\colhead{(mag)} & \colhead{(mag)} &
\colhead{($10^9L_\odot$)} & \colhead{(Mpc)} & 
\colhead{} & \colhead{}
}
\startdata
0033 & 12h11m07.76s & +14d16m29.8s & $20.8 \pm 4.9$ & $-16.39$ & $-17.01$ & $0.43\pm0.25$ & 1.48 & $0.61\pm1.17$ & $0.34\pm0.66$ \\
0230 & 12h17m19.64s & +11d56m36.2s & $27.3 \pm 1.5$\tna & $-16.21$ & $-16.96$ & $0.69\pm0.29$ & 0.96 & $9.39\pm2.19$ & $4.73\pm1.10$ \\
1087 & 12h28m14.90s & +11d47m24.0s & $42.0 \pm 1.5$ & $-17.79$ & $-18.64$ & $3.29\pm1.07$ & 0.25 & $5.07\pm0.73$ & $2.31\pm0.33$ \\
1185 & 12h29m23.43s & +12d27m02.4s & $29.5 \pm 0.8$\tna & $-16.77$ & $-17.37$ & $1.24\pm0.42$ & 0.10 & $2.73\pm1.11$ & $1.58\pm0.64$ \\
1355 & 12h31m20.04s & +14d06m53.5s & $20.3 \pm 4.7$ & $-17.51$ & $-18.16$ & $1.82\pm0.35$ & 0.50 & $1.07\pm0.56$ & $0.59\pm0.30$ \\
1407 & 12h32m02.69s & +11d53m24.8s & $31.9 \pm 2.1$ & $-16.72$ & $-17.43$ & $1.24\pm0.42$ & 0.17 & $10.18\pm1.76$& $5.30\pm0.92$ \\
1528 & 12h33m51.62s & +13d19m21.3s & $47.0 \pm 1.4$ & $-17.16$ & $-18.04$ & $1.63\pm0.48$ & 0.34 & $5.57\pm1.04$ & $2.46\pm0.46$ \\
1539 & 12h34m06.77s & +12d44m30.1s & $23.8 \pm 0.4$\tna & $-16.05$ & $-17.12$ & $0.52\pm0.11$ & 0.25 & $11.83\pm2.67$& $4.40\pm0.99$ \\
1545 & 12h34m11.54s & +12d02m55.9s & $38.0 \pm 1.0$\tna & $-16.91$ & $-17.74$ & $1.41\pm0.42$ & 0.26 & $9.37\pm1.52$ & $4.34\pm0.71$ \\
1695 & 12h36m54.87s & +12d31m12.5s & $24.4 \pm 2.2$ & $-17.49$ & $-18.32$ & $1.69\pm0.78$ & 0.43 & $1.45\pm0.58$ & $0.68\pm0.27$ \\
1895 & 12h41m52.00s & +09d24m10.3s & $23.8 \pm 3.0$ & $-16.60$ & $-17.29$ & $0.76\pm0.39$ & 1.16 & $1.45\pm0.99$ & $0.76\pm0.52$ \\
\enddata
\tablecomments{Coordinates are from \citet{Cote_04}. $\sigma_e$ for VCC~33, VCC~1087, VCC~1355, VCC~1407, VCC~1528, VCC~1695, VCC~1895 are measured by \citet{Toloba_14_II}. $\sigma_e$ for the remaining galaxies are from Keck/ESI spectra described in the text. Other parameters are from \citet{P08}. $S_{N,z}$ has the same definition as $S_N$, but normalized to $M_z=-15$.}
\tablenotetext{a}{Measured using Keck/ESI spectra.}
\end{deluxetable*}

\subsection{Observations}

\begin{figure}
\plotone{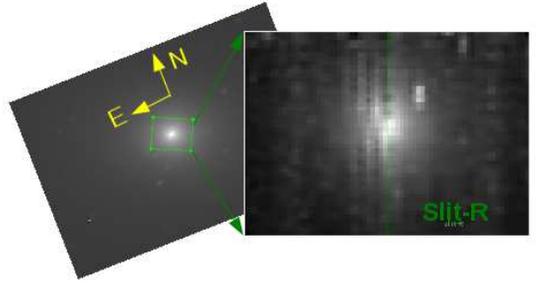}
\caption{Left: An image of VCC1545 taken by the Next Generation Virgo Cluster Survey \citep{Ferrarese_12}. The central region within the green region is the $7\arcsec \times 5\arcsec$ field of view of our GMOS-IFU pointing. The horizontal direction is along the major axis of the galaxy. Right: An image of VCC~1545 synthesized by stacking the 3D data cubes of our IFU data along the wavelength direction. The right side of the green vertical line is the region from slit-R, the slit used in our work. It's clear that the image of slit-B is not smooth as the "spaxels" inside are not linearly responding flux, so we have ignored this slit in our analyses. \label{IFU_image}}
\end{figure}

The observations were performed with the Integral Field Unit (IFU) \citep{A-S_02} equipped on the Gemini MultiObject Spectrograph (GMOS) \citep{Hook_04} at the Gemini South Telescope in Chile. We used the B600 grating with the $g'$ filter in two-slit mode (slit-B and slit-R). Each pseudo-slit contains 750 hexagonal elements (750 lenslet-coupled fibers) with a projected diameter of $0\farcs2$ each. A total of 1500 fibers are divided into 2 groups: 1000 are pointed to a contiguous object field of $7\arcsec \times 5\arcsec$, and the other 500 are pointed to a $5\arcsec \times 3.5\arcsec$ sky field, $1\arcmin$ away from the center of science field. Figure~\ref{IFU_image} shows the field of view superimposed on one of our sample galaxies, VCC~1545. Our sample galaxies have sizes in the range $9\arcsec < R_e < 31\arcsec$, where $R_e$ is the effective radius of a fitted Sersi\'c profile from \citet{Ferrarese_06}. 
Thus, our field-of-view only covers the central regions of our target galaxies.  

The spectra were binned by a factor of two in the wavelength direction, which reduces read noise and still oversamples the line spread function, delivering a dispersion of $0.9{\rm \AA/pixel}$. Each galaxy in our sample has four exposures. To reach $S/N=50 {\rm \AA}^{-1}$ at $5000{\rm \AA}$ ($S/N=67$ per binned pixel), the exposure time for individual galaxies varies from $4\times400$s to $4\times2000$s, depending on the surface brightness. Because the GMOS detectors consist of three CCDs, there  are two gaps in the wavelength direction. Therefore, we shift the central wavelength of two of the four exposures to dither across the CCD gaps.

Among our sample galaxies, VCC~1545 and VCC~1895 were observed in 2008A, and the others were observed in 2009A. In 2008, GMOS's middle CCD had a large amount of pattern noise that was not removed with bias subtraction, and also had many bad columns, which made the data calibration difficult. For the spectra from the left (B) slit, many important spectral features (H$_\beta$, Mgb, and the Fe indices) all fell on the middle CCD. Ultimately, we decided not to use slit-B data since the $S/N$ was still high enough when combining spectra from only slit-R. In order to maintain consistency across all galaxies in our sample in the calibration of stellar indices, we also chose to exclude the spectra from the left slit taken in 2009. 

Using the same instrument, we also observed five Lick standard stars (two G- and three K-type stars): HD114946, HD134439, HD149161, HD184406, and HD195633. As with the galaxies, each star was observed with two exposures, one slightly offset from the other in central wavelength to dither across the CCD gaps.

Although the line-of-sight distances are available from the surface brightness fluctuations (SBF) work of \citet{Mei_07} and \citet{Blakeslee_09}, we use projected cluster-centric distance in the following discussion because the SBF distances for our low-mass, low surface brightness ETGs have large error bars. Our results do not change if we use the 3-D distances.

\subsection{Data Reduction}
\label{DR}

We first removed cosmic rays from individual exposures using the L.A.Cosmic software package  (\citeauthor{vanDokkum_01_CR}, 2001). Subsequently, we used the GMOS-IFU IRAF pipeline to reduce the data, Because of our targets' low surface brightness, the level of the spectral continua are sensitive to the time-variable sky level, which means that the combined spectra can exhibit discontinuous jumps across the CCD gaps. Therefore, we decided to normalize the spectra, keeping track of their noise properties, before stacking them. We reduced all the standard stars in exactly the same way. 

All the galaxies in our sample contain nuclei (i.e., nuclear star clusters), some of which are bright and may have a different stellar population from the main body of the galaxy \citep{Cote_06,Paudel_10_nu}. Because the intrinsic sizes of nuclei are much smaller than the typical seeing of our observations ($0\farcs8$), the nuclei are essentially point sources. To exclude most of the nuclear light, we excluded all spectra inside the full width half maximum of a point source at the galaxy center when combining the galaxy spectra. This region is a small fraction of the total area covered by the IFU.

\begin{figure}
\epsscale{1.25}
\plotone{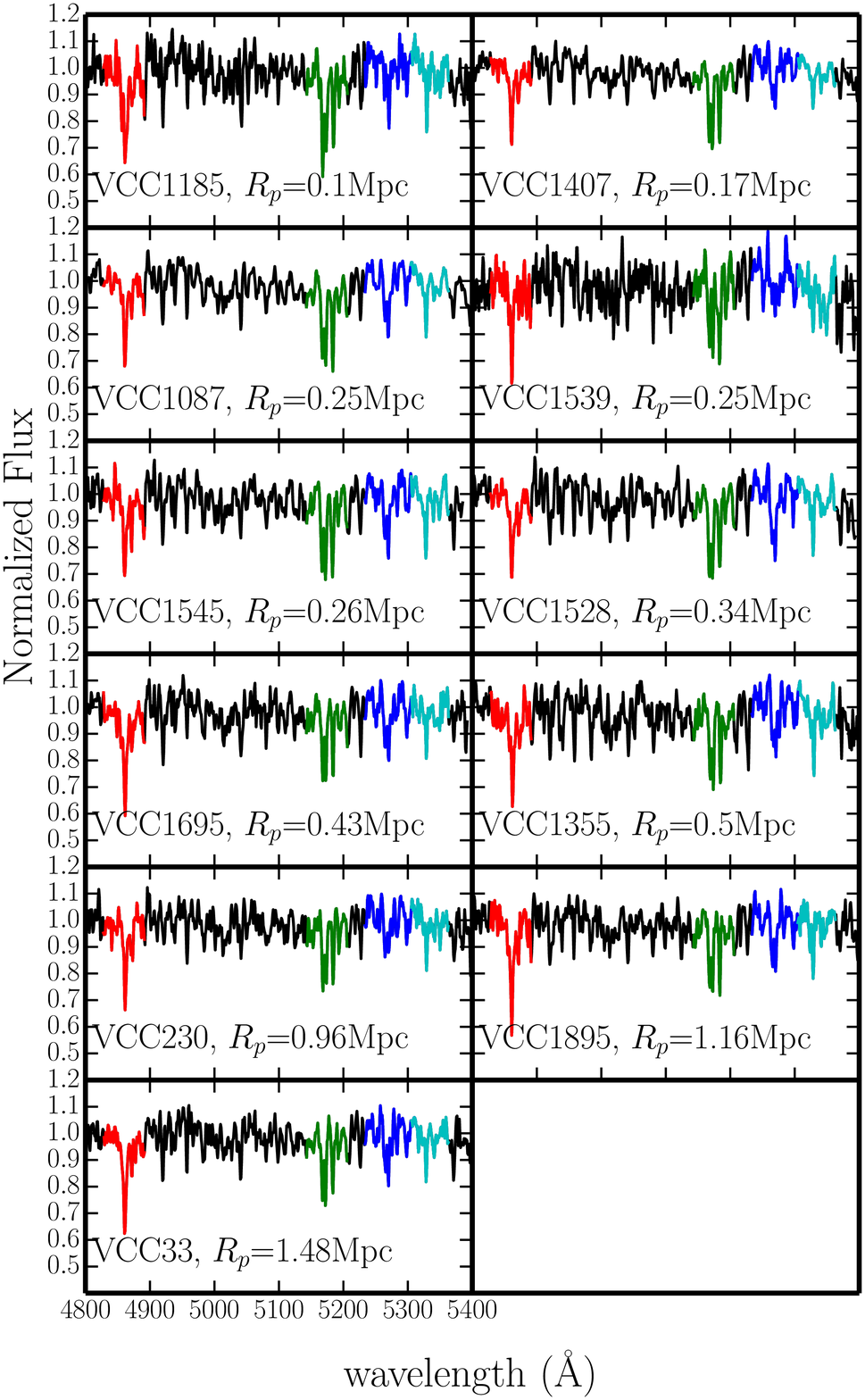}
\caption{The normalized spectra of each galaxy, excluding nuclei. Red, green, blue and cyan mark the wavelength regions for the H$_{\beta}$, Mgb, Fe5270, Fe5335 indices (absorption line and pseudo-continua), respectively. The plots are in order of increasing projected distance from M87 (left to right, then top to bottom). 
 \label{showspec}}
\end{figure}

Figure~\ref{showspec} presents the final spectra on which we perform our Lick index measurements. Red, green, blue and cyan mark the wavelength regions for the $H_{\beta}$, Mgb, Fe5270, Fe5335 indices from \citet{W94} (absorption lines and pseudo-continua), respectively. The 11 galaxies are plotted in order of increasing projected distance from M87. 

\section{Stellar Population Analysis}
\label{ssp}

\subsection{Lick Indices}

Lick indices were measured following the description in \citet{W94} (Eq.(1)-(3)). To compare with the Lick/IDS system, our medium-resolution spectra ($\sim3 {\rm \AA}$~FWHM at 5100{\rm \AA}, $\sigma\sim$75~km/s) were broadened to $\sim8.4-10{\rm \AA}$. The typical velocity dispersion of low mass ETGs is $\sim$50~km/s, which only changes the line width by about $0.1{\rm \AA}$, much less than the errors in the measurements, so we did not correct for the velocity dispersion. 
We used the variance arrays produced by our data reduction process to calculate the errors following the method of \citet{Cardiel_98}. We checked to see if these uncertainties are underestimated using a Monte Carlo approach, where we simulated the spectra using the known pixel variances and re-measured the Lick indices. The uncertainties we derived for the index measurements are similar to those derived using the analytic method.

For each index, we calibrated the measurements using the weighted mean offsets (between observed and published) for five observed Lick standard stars from \citet{W94}. 
These stars were chosen to span a range of effective temperature and metallicity that encompassed the expected range for our target galaxies. The $1\sigma$ uncertainties in the weighted mean difference between our measurements and the measurements of \citet{W94} are 0.058, 0.041, 0.046, and 0.035 for $H_\beta$, Mgb, Fe5270, and Fe5335, respectively. These systematic uncertainties are smaller than the typical random uncertainties in the index measurements. The calibrated indices for our target galaxies are listed in Table~\ref{I_dE}.

\begin{table*}
\begin{center}
\caption{Calibrated, integrated Lick index measurements \label{I_dE}}
\begin{tabular}{ccccccc}
\tableline\tableline
VCC & $H_\beta$ & [\ion{O}{3}]$_{5007\AA}$ & $Fe5015$ & $Mgb$ & $Fe5270$ & $Fe5335$ \\
\tableline\tableline
0033 & $2.42\pm0.07$ & $0.53\pm0.04$ & $4.34\pm0.12$ & $1.60\pm0.07$ & $1.63\pm0.08$ & $1.55\pm0.09$ \\
0230 & $1.89\pm0.10$ & $0.87\pm0.05$ & $3.74\pm0.15$ & $1.63\pm0.09$ & $1.90\pm0.10$ & $1.33\pm0.11$ \\
1087 & $1.82\pm0.05$ & $0.95\pm0.03$ & $4.61\pm0.08$ & $2.76\pm0.05$ & $2.37\pm0.05$ & $1.86\pm0.06$ \\
1185 & $2.46\pm0.15$ & $0.63\pm0.07$ & $4.26\pm0.23$ & $2.96\pm0.13$ & $1.87\pm0.14$ & $2.04\pm0.16$ \\
1355 & $2.24\pm0.12$ & $0.95\pm0.06$ & $5.84\pm0.18$ & $1.94\pm0.11$ & $1.77\pm0.12$ & $2.30\pm0.14$ \\  
1407 & $1.86\pm0.07$ & $0.67\pm0.03$ & $3.76\pm0.10$ & $2.45\pm0.06$ & $1.72\pm0.07$ & $1.30\pm0.08$ \\
1528 & $2.20\pm0.05$ & $0.94\pm0.03$ & $4.45\pm0.08$ & $2.93\pm0.05$ & $2.63\pm0.06$ & $1.86\pm0.07$ \\
1539 & $1.65\pm0.19$ & $0.85\pm0.09$ & $3.51\pm0.29$ & $2.86\pm0.16$ & $1.87\pm0.18$ & $2.40\pm0.22$ \\
1545 & $1.73\pm0.06$ & $0.98\pm0.03$ & $4.74\pm0.10$ & $2.68\pm0.06$ & $2.66\pm0.07$ & $2.04\pm0.08$ \\
1695 & $2.68\pm0.07$ & $0.68\pm0.03$ & $4.94\pm0.11$ & $1.62\pm0.07$ & $1.96\pm0.07$ & $1.63\pm0.09$ \\
1895 & $2.90\pm0.08$ & $0.69\pm0.04$ & $3.59\pm0.13$ & $1.66\pm0.08$ & $1.85\pm0.08$ & $1.46\pm0.10$ \\
\tableline
\end{tabular}
\end{center}
\end{table*}

\subsection{SSP model}

We compared our observations to the models of \citet[][hereafter TMB03]{TMB03}. 
TMB03 provides grids of age, metallicity, and [$\alpha$/Fe] corresponding to different Lick indices. To better visualize the data at low [$\alpha$/Fe], we linearly extrapolated [$\alpha$/Fe] to $-0.3$ and interpolated it into a finer grid with [0.1~Gyr, 0.01~dex, 0.01~dex] interval of [age, [Z/H], and [$\alpha$/Fe]]. We emphasize that we are most interested in relative measurements between our sample galaxies.

To maximize signal-to-noise, we chose indices redder than $\sim4500 {\rm \AA}$. Within the wavelength range of our observation, the best choices are $H_\beta$, Mgb, Fe5270, and Fe5335, which together are influenced by the age, $\alpha$-element abundance, and metallicity. 
For each galaxy, we found both the minimum $\chi^2$, and the reduced $\chi^2$ closest to 1, finding that both criteria give the same results. We then produce 1000 Gaussian-random distributed simulated observations based on the observed errors for every index, and we estimated the best-fit age, metallicity, and [$\alpha$/Fe] for each one. These results were used to determine the uncertainties at the 68\%  ($1\sigma$) confidence level. 
The fitting results of the SSP parameters are listed in Table~\ref{SSP_dE}.
All the error bars in the table are limited to the boundary of the model grids, i.e. a derived age plus its error bar cannot be older than the oldest model, and the same applies to metallicity and \afe. 

\subsection{Model Grids}

\begin{figure}
\epsscale{1.2}
\plotone{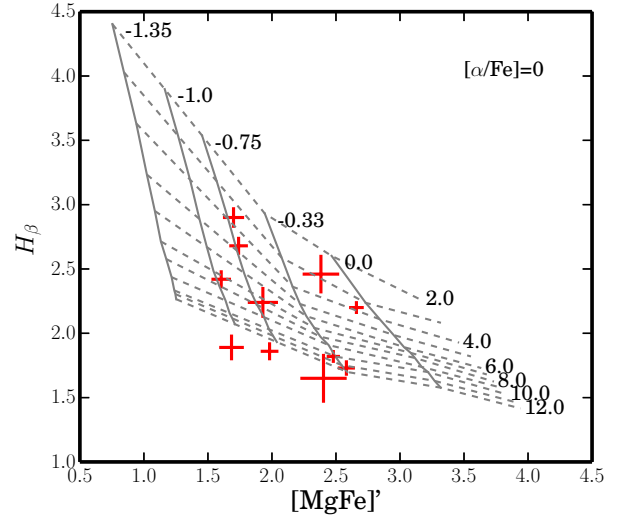}
\caption{ The age-[Z/H] grid from the TMB03 model at \afe=0. $[MgFe]'$ plotted along the x-axis, is chosen because it is relatively insensitive to \afe. Solid and dashed lines represent constant metallicity (dex) and ages (Gyr), and the numbers are given at the ends of each line. Eleven galaxies in our sample are displayed as red crosses with index error bars. Three of them, VCC~230, VCC~1407, and VCC~1539 are off-grid with low values of $H_\beta$, but VCC~1539 is not more than 1-$\sigma$ away, and the other two are within 2-$\sigma$ of the oldest ages. 
 \label{off-grid}}
\end{figure}

To give a more intuitive view, we plot these galaxies' index measurements on the TMB03 model grids.
In Figure~\ref{off-grid}, we plot $H_\beta$ versus $[MgFe]'$, where 
\begin{equation}
[MgFe]' = \sqrt{Mgb\times(0.72Fe5270+0.28Fe5335)}.
\end{equation}
Three galaxies have $H_\beta$ values off the model grids in the direction of old ages, but only two galaxies are more than $1\sigma$ away from the grid, and neither is $2\sigma$ away. We list their ages with an lower limit of 12.0~Gyr. Among them, VCC~1407 is in common with \citet{Toloba_14_II}, and it is off-grid in that study as well. 
Such off-grid phenomena are not uncommon in studies of old stellar populations, even for Milky Way GCs \citep[e.g.,][]{Poole_10}, and indicate that the models are not complete. 

One possibility is that weak $H_\beta$ absorption is due to low level of $H_\beta$ emission filling in the line. In this case, we would expect to see [\ion{O}{3}] emission at 5007\AA, and possible [\ion{N}{1}] emission, which might affect our Mgb measurement. To test for the presence of weak emission, we measured the [\ion{O}{3}]$_{5007}$ index defined by \citet{Gonzalez_93} (Table~\ref{I_dE}), for all 11 galaxies in our sample. None had a negative value (emission). Following \citet{Buzzoni_15}, we measured Fe5015 (corrected by Lick standard stars; in Table~\ref{I_dE}) and calculated the intrinsic [\ion{O}{3}] emission ([\ion{O}{3}]$'$), with similar results. Thus, we are fairly confident that weak emission is not affecting our line index measurements. As the [\ion{N}{1}] doublet is not expected to be observed without strong $H_\beta$ and [\ion{O}{3}] emission \citep{Sarzi_06}, our Mgb measurements should not be compromised. 

Another factor that can affect Balmer line strengths is the morphology of the horizontal branch (HB) at low metallicity in these galaxies. The models of TMB03 choose blue HBs below [Z/H]$\sim-1$, following the trend seen in Galactic GCs. Redder HBs, however, would make the grids move toward weaker H$_\beta$ at low metallicity. The \citet{MT00} models provide an option for red HBs at low metallicity, and although their work is based on solar-scaled abundances, we can use them to examine the effects of HB morphology in our metallicity range of interest. Assuming a red HB morphology, all our data are located within the model grids. Thus, a higher fraction of red HB stars than that in Galactic GCs at low metallicity is a plausible reason for why a few old galaxies are off the TMB03 model grid. 
Fortunately, the parameter space of metallicity and \afe, based on Mgb, Fe5270, and Fe5335, is almost independent of the age, and so our conclusions are not affected by this issue.

\begin{table}
\begin{center}
\caption{The SSP parameters of our sample galaxies. \label{SSP_dE}}
\begin{tabular}{cccr}
\tableline\tableline
VCC & $age (Gyr)$ & $[Z/H]$ & $[\alpha/Fe]$ \\
\tableline\tableline
0033 & $7.2^{+7.7}_{-0.9}$  & $-0.95^{+0.09}_{-0.15}$ & $-0.08^{+0.09}_{-0.15}$  \\
0230 & $>12.0$ & $-1.01^{+0.04}_{-0.03}$ & $-0.10^{+0.08}_{-0.05}$ \\
1087 & $10.7^{+4.1}_{-1.7}$ & $-0.37^{+0.07}_{-0.10}$ & $0.16^{+0.04}_{-0.03}$ \\
1185 & $3.2^{+1.4}_{-1.1}$  & $-0.16^{+0.21}_{-0.09}$ & $0.29^{+0.09}_{-0.14}$ \\
1355 & $7.3^{+6.7}_{-2.6}$  & $-0.65^{+0.17}_{-0.29}$ & $-0.13^{+0.11}_{-0.17}$ \\ 
1407 & $>12.0$ & $-0.75^{+0.02}_{-0.02}$ & $0.36^{+0.04}_{-0.03}$ \\
1528 & $4.4^{+0.6}_{-0.8}$  & $-0.13^{+0.05}_{-0.03}$ & $0.14^{+0.03}_{-0.03}$ \\
1539 & $>12.0$ & $-0.43^{+0.12}_{-0.14}$ & $0.17^{+0.14}_{-0.14}$ \\
1545 & $11.8^{+3.0}_{-1.9}$ & $-0.34^{+0.05}_{-0.10}$ & $0.03^{+0.05}_{-0.05}$ \\
1695 & $4.7^{+0.4}_{-1.0}$  & $-0.73^{+0.13}_{-0.06}$ & $-0.16^{+0.11}_{-0.07}$ \\
1895 & $3.7^{+0.8}_{-0.8}$  & $-0.69^{+0.11}_{-0.10}$ & $-0.04^{+0.10}_{-0.04}$ \\
\tableline
\end{tabular}
\tablecomments{The quoted uncertainties are the ranges encompassing 68\% of the Monte Carlo realizations.}
\end{center}
\end{table}

\begin{figure}
\epsscale{1.2}
\plotone{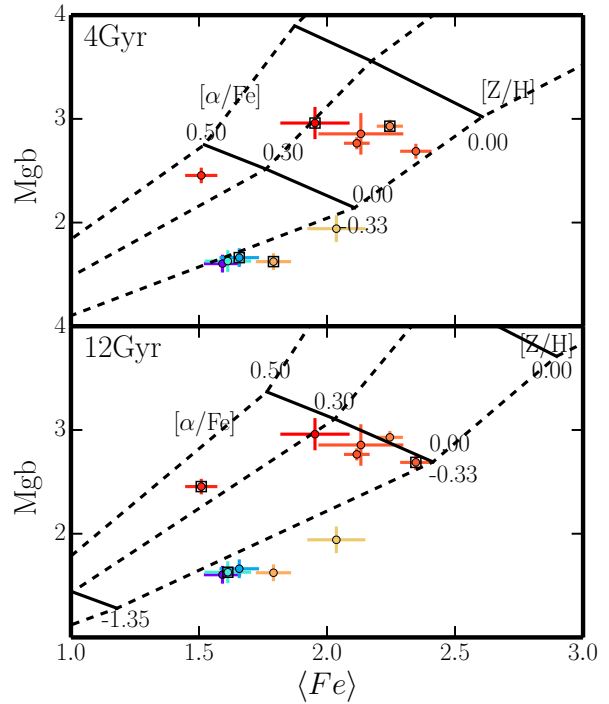}
\caption{The [Z/H]-[$\alpha$/Fe] grids of TMB03 model at 4 Gyr and 12 Gyr ages. The indices are Mgb and $\langle Fe \rangle$, which are not very sensitive to age variation. Both are metal lines, representing the abundance of $\alpha$-element and iron respectively. The 11 low mass ETGs are plotted with colors from red to blue, representing increasing projected distance to M87.  The ones in the inner region generally have higher [$\alpha$/Fe].
 \label{grid_zar}}
\end{figure}

To visualize trends in [Z/H] and \afe, we plot Mgb against $\langle Fe \rangle$ in Figure~\ref{grid_zar}, where $\langle Fe \rangle = (Fe5270 + Fe5335)/2$ is a combined index that is sensitive to iron abundance, but not very sensitive to age. Figure~\ref{grid_zar} shows our sample galaxies on the model grids for two different ages (4 and 12~Gyr) to show how age affects the estimate of [$\alpha$/Fe]. We note that the relative values of [$\alpha$/Fe] are not very sensitive to the estimated age. The SSP-equivalent age and metallicity are also determined in our fitting procedure. We represent projected cluster-centric distances by color, distance increasing from red to blue. In these plots, we can begin to see that the inner low mass ETGs generally have higher [$\alpha$/Fe] than the outer ones.

\section{Results}
\label{results}

\subsection{The \afe-$\sigma$ relation}

\begin{figure}
\epsscale{1.3}
\plotone{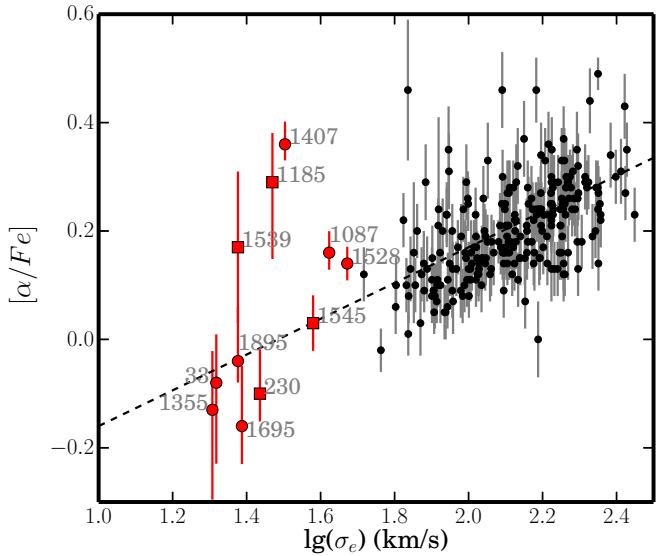}
\caption{ [$\alpha$/Fe]-$\sigma$ diagram for ETGs. Black circles are from ATLAS$^{3D}$ sample ([$\alpha$/Fe] and velocity dispersion measurements are from McDermid et al. 2015 and Cappellari et al. 2013 respectively). The black dash line is the [$\alpha$/Fe]-$\sigma$ relation of massive ETGs (fitted by black circles). Red circles and squares are the low mass ETGs which have velocity dispersion measured by Toloba et al. (2014) and C{\^o}t{\'e} respectively. 
The [$\alpha$/Fe]--$\sigma$ relation for massive galaxies doesn't work for low mass ones, for red symbols have larger scatter than 1$\sigma$ error of the relation. 
 \label{afe_sgm} }
\end{figure}

For more massive ETGs, there is a reasonably well-defined relation between [$\alpha$/Fe] and the velocity dispersion of the galaxy, where the more massive galaxies have higher [$\alpha$/Fe] (e.g., \citealt{Thomas_05,Annibali_07,McDermid_15}). In Figure~\ref{afe_sgm}, we reproduce the [$\alpha$/Fe]$-\sigma$ relation for 245 galaxies (black circles) in the ATLAS$^{3D}$ sample \citep{Cappellari_11}, with [$\alpha$/Fe] measurements from \cite{McDermid_15} and velocity dispersions from \cite{Cappellari_13}. 
The \afe\ we use here is measured within $R_e/2$ of galaxies, which is comparable to the scale on which we observe our sample galaxies.
\cite{McDermid_15} remark that the ATLAS$^{3D}$ galaxies do show a tendency, at fixed mass, to have higher [$\alpha$/Fe] in denser environments, but velocity dispersion (i.e., stellar mass) is still the main predictor of [$\alpha$/Fe]. The black solid line is the [$\alpha$/Fe]-$\sigma$ relation as fit to these massive ETGs.

What we find for our low mass ETGs is quite different. The [$\alpha$/Fe] values for our Virgo low mass ETGs have a large scatter at fixed $\sigma$. We show this in Figure~\ref{afe_sgm}, where we plot our sample alongside the more massive, ATLAS$^{3D}$ galaxies (black circles). Red circles and squares represent the seven low mass ETGs in our sample that have velocity dispersion measurements from \citet{Toloba_14_II} and our Keck observations, respectively. 

We note that the \afe\ values in the two samples were derived using different models, with \citet{McDermid_15} using a customized version of the \citet{Schiavon_07} models. Therefore, there may be some systematic uncertainties in comparing the two data sets. Our main point, however, is to show that the relative measurements of \afe\ in our low mass ETG sample shows a larger spread than what is seen in the sample of higher mass ETGs. Perhaps coincidentally, some of our low mass galaxies lie along the extrapolation of the relation derived for more massive ETGs. Some of the others, however, have [$\alpha$/Fe] values much higher than would be expected given their velocity dispersion. The low mass ETGs, in fact, have a spread in \afe\ values that span the full range of \afe\ for the ATLAS$^{3D}$ galaxies.

\subsection{\afe\ versus cluster-centric distance}

The large spread in \afe\ values for our low mass ETG sample suggests that there is another parameter controlling their behavior. Given that these are low mass cluster galaxies, environment is a natural candidate. 

\begin{figure}
\epsscale{1.3}
\plotone{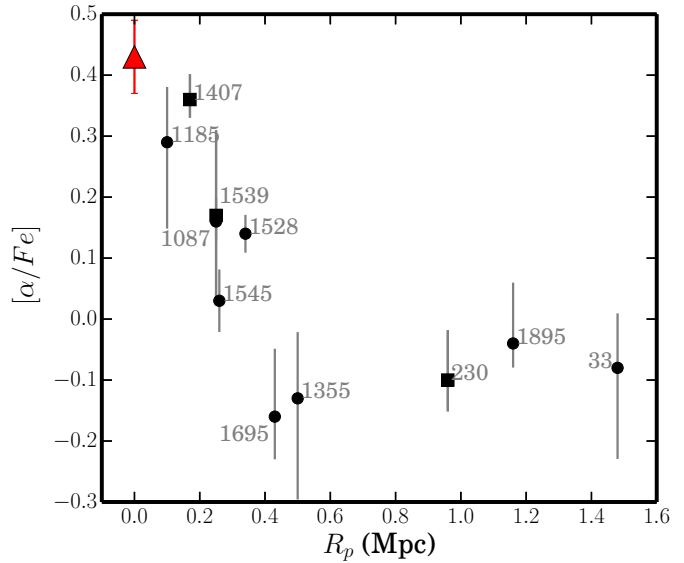}
\caption{The relation between SSP-equivalent [$\alpha$/Fe] and projected distance from M87. The red triangle represents M87 within R$_e$/2 with 1-$\sigma$ error bar, from the work of McDermid et al.\ (2015). Black circles are the low mass ETGs in our sample. The error bars represent $68\%$ confidence levels. The parameter [$\alpha$/Fe]  decreases with cluster-centric distance in the inner region, and then has a flat distribution in the outer regions, at values close to solar. 
The Spearman rank correlation coefficient is $-0.81\pm0.09$ for all these 11 galaxies, and it becomes $-0.92\pm0.11$ when only galaxies inside 0.6~Mpc are taking into account. 
\label{aFe_r}}
\end{figure}

Figure~\ref{aFe_r} shows the environmental dependence of [$\alpha$/Fe] by plotting \afe\ versus the projected distance from the center of the Virgo A subcluster, the cD galaxy, M87. Within the inner 0.4~Mpc (0.26$R_{vir}$, where $R_{vir}=1.55$~Mpc; \citealt{Ferrarese_12}), there is a gradient in [$\alpha$/Fe], where the low mass ETGs closer to the M87 have higher values. In the outer regions, $0.4<R_p<1.5$~Mpc, [$\alpha$/Fe] becomes flat and close to the solar value. 
For all 11 galaxies, the Spearman rank correlation coefficient is $-0.81$. If taking into account only the 8 galaxies inside 0.6~Mpc (6 inside 0.4~Mpc plus 2 slightly further), it goes to $-0.92$, which is at even higher significance. 
For reference, we also plot the [$\alpha$/Fe] value for the center of M87 at $R_p=0$ (McDermid et al.\ 2015). 

Our results show that the spread in \afe\ at low velocity dispersion can be explained as an environmental effect, where the low mass ETGs closest to the cluster center have the highest values. These values are similar to those seen in more massive ETGs.

\subsection{\afe\ and Globular Clusters}
The strong trend that we see in \afe\ versus cluster-centric distance (Figure~\ref{aFe_r}) supports the idea that local environment strongly influences the star formation histories of low mass Virgo ETGs. Although we cannot know the star formation histories of these galaxies in detail, the higher \afe\ values of the central ones tell us that they had shorter periods of star formation than the outer ones. For a similar amount of stellar mass produced, this implies that the central low mass ETGs had higher peak SFRs.

High star formation rates are believed to be one of the necessary conditions for forming massive star clusters (e.g., \citealt{Larsen_Richtler_99I,Weidner_04,Bastian_08}). \cite{P08} suggested that the central dwarf ETGs in Virgo must have had more intense star formation, because they had higher overall globular cluster specific frequencies. In fact, all the low mass galaxies in the ACSVCS with high $S_N$ were within the central 1~Mpc around M87. For our sample of eleven galaxies, we can now test if there is a correspondence between $S_N$ and \afe. 

\begin{figure}
\epsscale{1.3}
\plotone{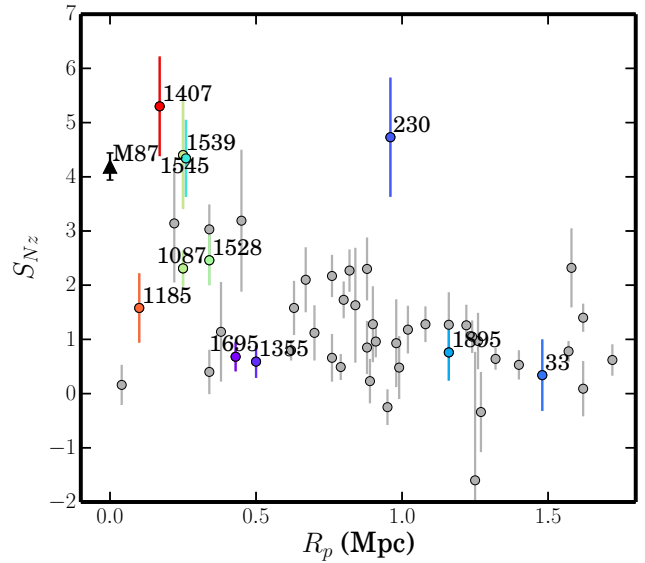}
\caption{Projected cluster-centric distance and $S_{N,z}$ for low mass ETGs with $M_z < -19$ in the ACS Virgo Cluster Survey. The grey circles are all the dwarf galaxies in the ACSVCS sample (from \citet{P08}), and the colored ones are from our sample. The colors represent the \afe\ measurements, from low to high (blue to red). The relatively high $S_{N,z}$ of M87 is added for comparison (black triangle). 
Note that the few negative $S_{N,z}$ values result from subtracting estimated numbers of GC contaminants from massive neighbors and background galaxies. $S_{N,z}\geq0$ for all galaxies, within the uncertainties. 
\label{Sn_p08}}
\end{figure}

\begin{figure}
\epsscale{1.3}
\plotone{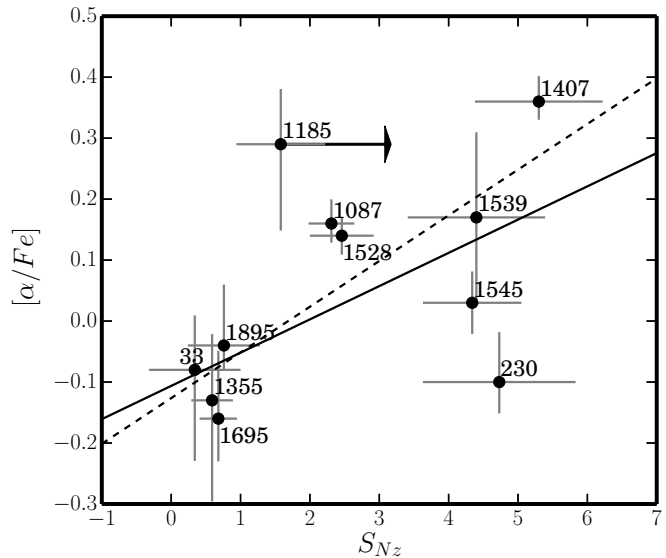}
\caption{SSP-derived $[\alpha/Fe]$ and $S_{N,z}$ ($S_N$ normalized in the $z\prime$-band). $[\alpha/Fe]$ and $S_{N,z}$ show a positive correlation. The solid line is fitted by excluding VCC1185, whose GC system may be in the process of being tidally stripped, and the dashed line is fit by excluding both VCC~1185 and VCC~230. The arrow on VCC~1185 represents our belief that the current $S_{N,z}$ value should be a lower limit.  
 \label{Sn_aFe}}
\end{figure}

Figure~\ref{Sn_p08} plots the relation between $S_{N,z}$ (specific frequency normalized to $M_z=-15$, as opposed to the usual $M_V=-15$) and distance from M87 for the ACSVCS sample of low mass ETGs (defined as $M_z>-19$). This plot shows the trend from \cite{P08}, where the innermost ones have the highest $S_{N,z}$. The eleven galaxies for which we have \afe\ measurements are labeled in color, with the color representing the \afe\ values (from blue to red as \afe\ increases). For reference, we also show the $S_{N,z}$ value for M87. In Figure~\ref{Sn_aFe}, we directly plot \afe\ versus $S_{N,z}$. 

In both Figures~\ref{Sn_p08} and \ref{Sn_aFe}, we can see a correlation between \afe\ and $S_{N,z}$. With two exceptions (VCC~1185 and VCC~230), the inner low mass ETGs that have high \afe\ also have higher GC specific frequencies. 
Taking into account all 11 galaxies, the Spearman rank correlation coefficient ($\rho_s$) is 0.56. To determine its significance, we performed bootstrapping resampling (5000 trials) where we calculated $\rho_s$ after randomly scrambling the association of \afe\ with $S_{N,z}$ in each trial. In the end, only 170 out of 5000 trials had $\rho\geq 0.56$. Our measured $\rho_s$ for the actual data is thus larger than $96.6\%$ of random, scrambled realizations of the data. 
When using the Pearson correlation coefficient ($\rho_p$), which measures linear correlations, it is 0.49 and larger than $86.5\%$ of all scrambled realizations. 
We contend that the measured values of $\rho_s$ and $\rho_p$ imply a correlation between \afe\ and $S_{N,z}$, supporting the idea that rapid star formation is associated with a higher mass fraction of globular clusters. 

One exception, VCC~1185, is the innermost galaxy in our sample. It has a relatively high \afe\ measurement of 
$0.29^{+0.09}_{-0.14}$~dex, but its $S_{N,z}$ is relatively low compared to other low mass ETGs at this distance from M87. Inspection of the distribution of GCs around VCC~1185 from the ACSVCS imaging shows that it is much more extended than other, comparable dwarfs \citep{P08}. We suspect that VCC~1185, being only $R_p=100$~kpc from M87, is already in the process of having its GCs stripped by the larger galaxy. Low mass galaxies in the ACSVCS that are closer to M87 (VCC~1327 and 1297) have $S_{N,z}$ values consistent with zero, and the likely reason is tidal stripping. Although we cannot prove that some of VCC 1185's GCs have been stripped, its apparent proximity to M87 makes it more possible than for other galaxies in our sample, which is why we consider its measured $S_{N,z}$ as a lower limit. When excluding VCC~1185 from the correlation analysis described above, we measure $\rho_s=0.62$ and $\rho_p=0.64$, which is larger than $97.5\%$ and $95.0\%$ of the random, scrambled trials respectively. 

The other exception, VCC~230, is more difficult to understand. Perhaps the one inconsistent fact about VCC~230 is its high $S_{N,z}$. It is nearly 1~Mpc away from M87 in projected distance, and does not lie close to any other massive galaxies in the cluster. Its low \afe\ value is consistent with its mass and environment. It's high $S_{N,z}$, however, is fairly certain. Its GCs are centrally concentrated around the galaxy, and there is little chance that they are background objects. Its luminosity-weighted age and metallicity are also not peculiar (12~Gyr and [Fe/H]$=-1.0$). One possible clue is that its GC luminosity function is fairly narrow, with a Gaussian sigma in the $g$-band of $\sigma_g=0.55\pm0.20$ \citep{Jordan_07}. A narrow GCLF translates into a steeper mass function, which suggests fewer massive GCs. If this lack of the most massive GCs is due to formation, and not disruption or inspiral due to dynamical friction, then this might be a hint that VCC~230's peak star formation rate was not quite as high as in other galaxies, despite still forming a large total number of star clusters. 

The linear fits shown in Figure~\ref{Sn_aFe} are for the full sample (solid) and the sample excluding VCC 1185 and 230 (dotted). Although these data are suggestive of a correlation between these two quantities, more data are needed to more firmly establish the relationship.

\section{Discussion}
\label{discussion}

\subsection{The Formation of Low Mass Cluster Galaxies}
\label{dE formation}

In this work, we have shown that the [$\alpha$/Fe] of low-mass early-type Virgo galaxies depends strongly on their distance from the cluster center. This integrated [$\alpha$/Fe] is a rough clock that measures the luminosity-weighted timescale for star formation in the galaxy, with higher [$\alpha$/Fe] values being due to a shorter duration (higher intensity) of star formation. The trend that we see, where the innermost low mass ETGs have higher [$\alpha$/Fe] values is consistent with the idea that galaxy evolution happens more quickly in denser environments. These galaxies may have fallen into the cluster halo earlier in their evolution \citep{Lisker_13}, and had their gas reservoirs depleted or removed on faster timescales. Other galaxy clusters have also shown hints of [$\alpha$/Fe] gradients in low mass ETGs (e.g., Abell~496 and the Coma cluster, \citealt{Chilingarian_08} and \citealt{Smith_09}), showing that the process is likely not unique to Virgo. Supporting evidence also comes from higher redshift observations. \citet{Santos_15} found an inverse SFR-density relation in a massive galaxy cluster at 
$z = 1.58$, where the star formation rate in the cluster core is four times higher than in the outskirts. A similar reversed SFR-density relation is also found in the field at $z\sim1$ \citep{Elbaz07} and in a galaxy group at $z = 1.62$ \citep{Tran_10}, a trend that is reproduced by simulations \citep{Tonnesen_Cen_14}. These studies show that environmental effects directly influence the early star formation process, and not only through quenching.

\subsection{The Build-up of Massive Galaxies}
\label{halo building}

In the standard hierarchical model of galaxy assembly, massive galaxies are built up by the accretion of satellite galaxies, particularly in their outer regions. Current ``two-phase'' models \citep{Naab_09,Oser_10} echo the early ideas on the formation of the Galactic halo \citep{Searle_Zinn_78,CMW98}. Observations of massive, compact high-redshift galaxies that have no counterpart in the local Universe require a significant amount of size growth that is currently best explained by the accretion of lower mass systems. One of the challenges of understanding this model is that most low mass galaxies studied to date have \afe\ values different from those in the stellar halos of massive galaxies. \citet{Greene_12,Greene_13} measured the mean ages, [Fe/H], and \afe\ for the outer halos of 25 massive ETGs, finding that these stellar populations had old ages, relatively low metallicity ([Fe/H]$\sim-0.5$), and high \afe$\sim+0.3$, a combination that had not been found in the low mass galaxies that could plausibly form an accreted population. As the relationship between \afe\ and galaxy velocity dispersion shows (Figure~\ref{afe_sgm}), simple dry merging of lower mass systems along this relation cannot reproduce the \afe\ values of more massive galaxies. 

\citet{Greene_13} suggested that stars in the outer halos of massive ETGs originated in low mass systems whose star formation histories were truncated around $z\sim2$. The innermost dwarf ETGs around M87 could be a remnant of this population, as they have [Fe/H] and \afe\ values that are similar to those seen in the outer halos of massive ETGs. Figure~\ref{aFe_r} compares our results (black circles) with the SSP-equivalent [$\alpha$/Fe] of M87 (red triangle; central R$_e$/2) from McDermid et al. (2015). As the cluster-centric distance decreases, the [$\alpha$/Fe] of low mass ETGs increases, and nearly meets the value for M87 itself. While these kinds of low mass galaxies cannot themselves build up all of M87 stars (their metallicities are too low, for instance), galaxies like them may have been important contributors to the outer halo stars and globular clusters. Although our data do not require that the M87 halo be made up of galaxies like the innermost dwarfs, the identification of this population of old, low-[Fe/H], high-\afe\ low mass ETGs in the vicinity of M87 suggests that this is a possibility. Moreover, the continuous and clear trend in \afe\ indicates a correlation between environment and star formation timescale, independent of mass. In these very dense regions, at least, environment seems to be the dominant factor controlling global star formation timescales.

\subsection{The Formation of Globular Clusters}
\label{GC}
We also present, for the first time, a correlation between the stellar \afe\ (i.e., formation timescale) and the specific frequency of globular clusters (i.e., mass fraction in massive star clusters). This points the way toward understanding the mechanisms underpinning the varying mass fraction of GCs across a wide range of galaxies. The question of why different galaxies have different $S_N$ values has been an unsolved puzzle since the quantity itself was defined \citep{Harris_vdB_81,Harris_01}. Over fifteen years ago, we started to understand that the GC mass fraction was roughly constant across galaxies when taken as a fraction of either the total baryonic mass \citep{McLaughlin_99} or total mass \citep{BTM97,Blakeslee_99}. This work was later refined across a wider range of galaxies using different ways to estimate galaxy total mass, such as halo occupation modeling (e.g., \citealt{P08,Spitler_Forbes_09,Hudson_14}), or stellar dynamics \citep{Harris_13}. 

Although a simple scaling between the number of GCs and the total mass of the system is both aesthetically appealing and provides a first-order description of the existing data, we still do not understand {\it why} GCs should have such a scaling when the other stars in a galaxy do not. 
Current thinking holds that GCs simply form at a fixed fraction of total gas mass, while the formation of the rest of the stars in a galaxy is subject to the various feedback processes that produce a mass-dependent star formation efficiency. The question of why the formation of GCs should not be subject to these feedback processes, however, is still an open one. One idea is that the GCs form early in the star formation process \citep{BTM97,Harris_Harris_02,P08}, making them little affected by the quenching process. The result is that stronger quenching, which decreases the overall SF efficiency, leads to higher specific frequencies \citep{Mistani_15}. Another possibility is that two galaxies can have similar SF efficiencies, but that the one with higher GC specific frequency had a star formation history more conducive to the production of massive star clusters, e.g., higher peak SFR \citep{Bastian_08} or higher SFR density \citep{Larsen_Richtler_00III}. Yet another idea is that GC mass fraction is telling us more about GC survival rather than formation. \citet{Kruijssen_14} present a model where the dynamics of mergers remove young massive clusters from dynamically hostile star forming disks, and deposit them into the relatively benign halos of galaxies, allowing them to survive. 

Our results suggest that low mass galaxies with high GC specific frequencies formed the bulk of their stellar mass over a relatively short period of time. We still cannot say, however, whether this is due to internal or external processes. Not knowing the total masses of our sample galaxies, we are not sure whether a constant GC mass fraction can explain the observed $S_N$. It is possible that the inner ones have lower field star formation efficiencies, perhaps due to environmentally-induced processes like ram pressure stripping, resulting in higher $S_N$. It is also possible that these central low mass ETGs had a star formation history that was more efficient at producing massive star clusters. It is likely that some combination of these two effects is involved. If merging is important for the formation \citep{Li_Gnedin_14} or survival \citep{Kruijssen_14} of GCs then perhaps merging also results in faster quenching times that would boost \afe. However the GCs are produced, the high-$S_N$, high-\afe\ stellar populations of the low-mass ETGs in the Virgo cluster core are quite similar to the stellar halos of massive ETGs, indicating a likely connection.


\section{Conclusions}
\label{conclusion}

We present integrated non-nuclear spectra and \afe\ measurements for 11 low mass ETGs in the Virgo Cluster at a range of distances from the cD galaxy, M87. Our sample was also chosen to have a range of globular cluster $S_N$. Our main conclusions are:

\begin{itemize}

\item We find a clear environmental dependence for \afe, with low mass early-type galaxies closer to M87 having higher \afe\ values. Within the innermost 0.4~Mpc (projected), low mass ETGs in our sample have super-solar \afe, while those in the outer regions have values around solar.

\item The inward extrapolation of the \afe\ gradient with distance matches the \afe\ value for M87 itself. This suggests that star formation timescales in the Virgo core are regulated by environment, and that our high-\afe\ galaxies may be transition objects between those farther out and the ones that have already been accreted into the M87 halo.

\item The high \afe\ galaxies in our sample galaxies do not lie on the \afe--$\sigma$ relation, indicating that the \afe\ of low mass ETGs in the cluster environment are not mainly driven by their mass. 

\item The [$\alpha$/Fe] values of low mass ETGs correlate with globular cluster $S_N$. This suggests that the high-$S_N$ galaxies had either a higher early star formation rate (or intensity), or a shorter timescale of star formation (i.e., faster quenching in the epoch of field star formation).

\end{itemize}

Our results imply that low mass ETGs in dense environments were formed (and perhaps also quenched) rapidly, and that they were efficient at forming, or retaining, massive star clusters. These galaxies are likely most similar to the objects accreted into the M87 halo. Future chemical abundance and dynamical mass measurements in these low mass ETGs will be helpful for determining the formation histories of these galaxies and their globular clusters.


\acknowledgements

YQL thanks the Gemini Observatory Helpdesk for their kind and helpful assistance with data reduction. She also thanks Jun Hou for useful discussions on stellar population models.
YQL and EWP acknowledge support from the National Natural Science Foundation of China under Grant Nos.\ 11173003 and 11573002, and from the Strategic Priority Research Program, ``The Emergence of Cosmological Structures'', of the Chinese Academy of Sciences, Grant No. XDB09000105. THP acknowledges support through FONDECYT Regular Project Grant No. 1121005. Both AJ and THP acknowledge support from BASAL Center for Astrophysics and Associated Technologies (PFB-06).
HXZ acknowledges support from China Postdoctoral Science Foundation under Grant No. 552101480582. HXZ also acknowledges support from CAS-CONICYT Postdoctoral Fellowship, administered by the Chinese Academy of Sciences South America Center for Astronomy (CASSACA). 

Based on observations obtained at the Gemini Observatory (acquired through the Gemini Science Archive and processed using the Gemini IRAF package), which is operated by the 
Association of Universities for Research in Astronomy, Inc., under a cooperative agreement 
with the NSF on behalf of the Gemini partnership: the National Science Foundation 
(United States), the National Research Council (Canada), CONICYT (Chile), the Australian 
Research Council (Australia), Minist\'{e}rio da Ci\^{e}ncia, Tecnologia e Inova\c{c}\~{a}o 
(Brazil) and Ministerio de Ciencia, Tecnolog\'{i}a e Innovaci\'{o}n Productiva (Argentina). Data for this paper were obtained through program numbers: GS-2008A-Q-33 and GS-2009A-Q-11.

Some of the data presented herein were obtained at the W.M. Keck Observatory, which is operated as a scientific partnership among the California Institute of Technology, the University of California and the National Aeronautics and Space Administration. The Observatory was made possible by the generous financial support of the W.M. Keck Foundation. The authors wish to recognize and acknowledge the very significant cultural role and reverence that the summit of Maunakea has always had within the indigenous Hawaiian community.  We are most fortunate to have the opportunity to conduct observations from this mountain.


\bibliographystyle{apj}
\bibliography{ref_VdEspIFU}

\clearpage 

\end{document}